\documentstyle[preprint,aps]{revtex}
\begin{document}
\draft
\title{Nuclear Density-Dependent Effective Coupling Constants in
the Mean-Field Theory}
\author{Jae Hwang Lee and Young Jae Lee}
\address{Department of Physics, Dankook University, Cheonan
330-714, Korea}
\author{Suk-Joon Lee}
\address{Department of Physics and Institute of Natural Sciences,
Kyung Hee University, Suwon, Korea}
\date{\today}
\maketitle

\begin{abstract}
It is shown that the equation of state of nuclear matter can be
determined within the mean-field theory of $\sigma \omega$ model
provided only that the nucleon effective mass curve is given.
We use a family of the possible nucleon effective mass curves that
reproduce the empirical saturation point in the calculation of the
nuclear binding energy curves in order to obtain density-dependent
effective coupling constants.
The resulting density-dependent coupling constants may be used to
study a possible equation of state of nuclear system at high density
or neutron matter.
Within the constraints used in this paper to $M^*$ of nuclear matter
at saturation point and zero density, neutron matter of large
incompressibility is strongly bound at high density while soft 
neutron matter is weakly bound at low density.
The study also exhibits the importance of surface vibration modes
in the study of nuclear equation of state.
\end{abstract}
\pacs{}

\section{INTRODUCTION}
The relativistic $\sigma\omega$ model in the mean-field theory
has been very successful in describing the properties of nuclear
matter and finite nuclei\cite{Wal74,W&S86,H&S87,Se79,H&S84}.
It provides a useful framework to evaluate the one-baryon-loop
vacuum fluctuation effects at finite density. In the mean-field
approximation one can obtain the baryon self-energies in nuclear
matter taking into account the Dirac sea modified due to the
presence of valence nucleons. The energy density contains the
modified negative energies, the correction that removes half
of the interaction energy between the states in the Dirac sea,
and the interaction energy between the valence particles and
those in negative energy states.

Recently, Brockmann and Toki\cite{B&T92} reported a relativistic
density-dependent Hartree calculation for finite nuclei, where
the results agree very well with experiment. They employ $\sigma$
and $\omega$ exchange for the effective interaction without taking
into account the vacuum fluctuation effects and adjust the
density-dependent effective coupling constants such that
the relativistic Hartree calculation reproduces the nuclear matter
Dirac-Brueckner-Hartree-Fock (DBHF) results based on a realistic
$NN$ interaction.
The density-dependent coupling constants may effectively contain
exchange term and contributions of iterated meson exchange. The
effective scalar field for instance, contains large contribution
from two pion exchange and the iterated exchange of $\pi$- and
$\rho$-mesons contribute to the effective vector fields.

We show that the energy per nucleon curve can be 
calculated in nuclear matter if the nucleon effective mass is 
given in the relativistic Hartree approach(RHA) with the 
inclusion of the vacuum fluctuation effects.
A brief summary about RHA with density-dependent effective 
coupling constants is given in Sections II and III.
A resonable family of the nucleon
effective mass curves is used to collect the density-dependent
coupling constant curves that reproduce saturation properties
of nuclear matter. These curves define the upper and lower
bounds of the density-dependent coupling constants in RHA where
one-baryon-loop vacuum effects are taken into account.
The resulting range of the density-dependent coupling constant
would restrict a possible equation of state of a neutron matter
or a nuclear system at high density.
It also would limit a possible nuclear incompressibility at low
density such as at the surface of finite nuclei which
is important in study of giant resonances related with
surface vibration.

\section{THE APPLICATION OF THE THERMODYNAMIC LAW IN NUCLEAR
MATTER}
The nucleon self-energy in nuclear matter has the general form
by assuming parity conservation, time-reversal invariance, and
hermiticity\cite{W&S86}:
\begin{equation}
\Sigma(k) = \Sigma^s(\underline{k},k^0,k_F)
- \gamma^0 \Sigma^0(\underline{k},k^0,k_F)
+ \bbox{\gamma}\cdot \bbox{k}\Sigma^v(\underline{k},k^0,k_F),
\end{equation}
where $ k = (k^0,\bbox{\rm k})$ is the nucleon four momentum
and $\gamma^\mu$ is Dirac matrix.
$\it k_F$ is Fermi momentum of nucleon in nuclear matter and
$\underline{k} = |\bbox{\rm k}|$, so that $0< \underline{k} \leq k_F$.
The self-energy can be reduced within the Dirac equation 
to a potential
\begin{equation}
U(\underline{k},k_F ) = A(\underline{k},k_F ) +
\gamma^0 B(\underline{k},k_F)
\label{self}
\end{equation}                                           
where the scalar potential $A$ and the vector potential $B$ are given by
\begin{eqnarray}
A(\underline{k},k_F )={{\Sigma^s -M\Sigma^v}\over{1+\Sigma^v}} \\
B(\underline{k},k_F )={{E\Sigma^v -\Sigma^0}\over{1+\Sigma^v}}.
\label{B}
\end{eqnarray}                                           
Here, $M$ is the free nucleon mass and $E$ is the single particle
energy of a nucleon in nuclear matter including the rest mass energy
of a free nucleon.
In a mean field approximation, $A$ is independent of 
$\underline{k}$ and define the nucleon effective mass 
$M^*(\it k_F)$ as
\begin{equation}
M^* (k_F ) = M + A(k_F).
\end{equation}
The nucleon 
in nuclear matter behaves like a free particle
with energy $E^*$ and mass $M^*$, where the relation
\begin{equation}
E^* = E - B = {{E+\Sigma^0} \over {1+\Sigma^v}}
= \sqrt{ {\underline{k}}^2 +{M^*}^2 }
\label{ESTAR}
\end{equation}
holds in the Dirac equation.

In the thermodynamic treatment, we shall regard nuclear matter
as the system containing the same kind of particles, namely
single-component system. Then, the fundamental thermodynamic law
is given by
\begin{equation}
dE_{int} = TdS-PdV+\mu dN,
\end{equation}
which describes the change in the internal energy $E_{int}$ arising
from small change in the entropy $S$, the volume $V$, and the
number of nucleons $N$ in $V$. We confine ourselves to
the ground state nuclear matter here, so the temperature $T$ is
zero. Furthermore, the Hugenholtz-Van Hove (HV) theorem states
that the chemical potential $\mu$ equals the highest occupied single
particle energy\cite{HV}. Therefore, we have
\begin{equation}
{P \over \rho_B} + E_A = e(M^* , k_F ),
\label{PPP}
\end{equation}
where $e(M^*,\underline{k}) \equiv E-M$ is the single particle energy,
$\rho_B = dN/dV$ the baryon density and $E_A=dE_{int}/dN$
the energy per nucleon.

Using the relation
\begin{equation}
P={\rho_B}^2 {dE_A \over d\rho_B}
\label{P1}
\end{equation}
and
\begin{equation}
\rho_B = {\gamma \over 6\pi^2} {k_F}^3
\end{equation}
where $\gamma$ is the spin-isospin degeneracy, it immediately 
follows from Eq. (\ref{PPP}) that
\begin{equation}
E_A + {k_F \over 3}{ dE_A \over dk_F} = e(M^* , k_F ).
\label{E1}
\end{equation}

On the other hand, from Eqs.(\ref{B}) and (\ref{ESTAR}) the single
particle energy $e(M^*,\underline{k})$ can be expressed as
\begin{equation}                                   
e(M^*,\underline{k}) \equiv E-M = 
\sqrt{ {\underline{k}^2}+{M^*}^2 }+B-M.
\label{single}
\end{equation}
Substituting Eq. (\ref{single}) for $\underline{k} = k_F$
into Eq. (\ref{E1}), we have
\begin{equation}
E_A + {k_F \over 3}{dE_A \over dk_F} = \sqrt{{k_F}^2+{M^*}^2 }+B-M .
\label{thermo}
\end{equation}

\section{THE NUCLEON BINDING ENERGY IN NUCLEAR MATTER}
The role of quantum corrections is crucial in the application of
hadronic fields theories to nuclear matter. We will use the 
relativistic Hartree approximation(RHA) which includes
one-baryon-loop vacuum effects. The RHA involves divergent integrals
over the occupied negative energy states. The divergences can be
rendered finite by including the appropriate counterterms in the
MFT Largrangian and by defining a set of renormalization conditions.
The vacuum fluctuation corrections to the binding energy per nucleon
can be obtained by carrying out the renormalization procedure. The
result of the binding energy per nucleon with the vacuum corrections
is given by\cite{W&S86}
\begin{equation}
E_A =   {B \over 2} - M
      + { m_s^2 \over {2g_s^2} }{ (M-M^*)^2 \over \rho_B }
      + { \gamma \over (2\pi)^3 \rho_B }
        \int_0^{k_F} d^3k \sqrt{ \underline{k}^2 +{M^*}^2 }
      + \Delta E_{VF},
\label{EAA}
\end{equation}
where
\begin{eqnarray}
\Delta E_{VF} = -{3\over{8k_F^3}} \left[
                   {M^*}^4 \ln{ \left( M^*\over M \right) }
                 + M^3(M-M^*)
                 - {7\over 2} M^2(M-M^*)^2
\right. \nonumber \\
\left.
                 + {13\over 3} M(M-M^*)^3
                 - {25\over 12} (M-M^*)^4   \right].
\end{eqnarray}
No counterterm for vector self-energy to be finite is
needed and therefore the RHA result for the vector self-energy
term is identical to that of the MFT which is given by
\begin{equation}
B = {g_v^2 \over m_v^2} \rho_B.
\label{VP}
\end{equation}
For given coupling constants, $(g_s/m_s)^2$ and $(g_v/m_v)^2$ as
functions of $k_F$,
Eq.(\ref{EAA}) is a function of the effective mass $M^*$ and
the Fermi momentum $k_F$.
The nuclear matter saturation is determined
by minimizing $E_A$ in $k_F$--$M^*$ space.

For each value of $k_F$, $E_A$ given by Eq. (\ref{EAA}) should take a
minimum at the right value of $M^*$, so that the partial derivative
of $E_A$ with respect to $M^*$ vanishes:
\begin{eqnarray}
M&-&M^* = {g_s^2 \over m_s^2}{\gamma \over (2\pi)^3}
        \int_0^{k_F} d^3k
          { M^* \over \sqrt{\underline{k}^2+{M^*}^2} }  \nonumber \\
       &-&{g_s^2 \over m_s^2}{1\over \pi^2}{\gamma \over 4}
        \left[ {M^*}^3 \ln{ \left( M^* \over M\right) }
                  -M^2(M^*-M)
                  -{5\over 2} M(M^*-M)^2
                  -{11\over 6} (M^*-M)^3
           \right].
\label{MMM}
\end{eqnarray}
By using Eq. (\ref{MMM}) the factor $m_s^2/g_s^2$ can be eliminated
in Eq. (\ref{EAA}) to have the expression of $E_A$ as
\begin{eqnarray}
E_A &&= {B\over 2} - M
       + {3\over 4 k_F^3} \left[ 
          \left( k_F^3+k_FMM^* - {1\over2}k_F{M^*}^2 \right)
          \sqrt{ k_F^2 + {M^*}^2 }
\right. \nonumber \\
      &&+\left( {1\over 2}{M^*}^4 - M{M^*}^3 \right)
          \ln{ \left({k_F + \sqrt{k_F^2+{M^*}^2}} \over M^* \right) }
        + \left \{
                    \left( {1\over 2}{M^*}^4 - M{M^*}^3 \right)
                    \ln{ \left( M^* \over M \right) } \right.
\nonumber \\
&&\left. \left.
        -{1\over 2} M^3(M-M^*)
        + {3\over 4} M^2(M-M^*)^2
        + {1\over 3} M(M-M^*)^3
        - {19\over 24} (M-M^*)^4  \right \}
       \right].
\label{EEE}
\end{eqnarray}
We combine Eq. (\ref{EEE}) with Eq. (\ref{thermo}) to eliminate
$B$ (i.e., $g_v^2/m_v^2$) and obtain the following relation:
\begin{eqnarray}
{d\over dk_F}&& \left( E_A \over k_F^3 \right) =
         { {3M+3\sqrt{k_F^2+{M^*}^2}}\over k_F^4 }
       - {9\over 2 k_F^7} \left[ 
            \left( k_F^3+k_FMM^*-{1\over 2}k_F{M^*}^2 \right)
            \sqrt{ k_F^2+{M^*}^2 }
\right. \nonumber \\
        &&+ \left( {1\over 2}{M^*}^4 - M{M^*}^3 \right)
            \ln{ \left( {k_F+\sqrt{k_F^2+{M^*}^2}}\over M^* \right) }
          + \left\{
                \left( {1\over 2}{M^*}^4 - M{M^*}^3 \right)
                \ln{ \left( M^* \over M \right) } \right.
\nonumber \\
&&\left. \left.
          - {1\over 2}M^3 (M-M^*)
          + {3\over 4}M^2 (M-M^*)^2
          + {1\over 3}M (M-M^*)^3
          - {19\over 24} (M-M^*)^4  \right\}
         \right]
\label{AAA}
\end{eqnarray}
One can solve Eq. (\ref{AAA}) numerically for $M^*(k_0)$ for a
given value of $E_A(k_0)=-15.75 \rm ~MeV$ by using the fact that
$dE_A/dk_F = 0$ at $k_F=k_0$ where $k_0$ is the saturation density
given by $1.42 \rm ~fm^{-1}$.

\section{CALCULATION OF DENSITY-DEPENDENT COUPLING CONSTANTS}

The scalar meson coupling constant and the nucleon effective mass
are related to each other through Eq. (\ref{MMM}) with $\gamma=4$.
The density-dependent coupling constant curve for the scalar
meson can be obtained immediately when $M^*(k_F)$ is given:
\begin{equation}
{g_s^2 \over m_s^2} = {{\pi^2(M-M^*)} \over F(k_F,M^*)}
\label{GS}
\end{equation}
where
\begin{eqnarray}
F(k_F,M^*) =  \left[ M^*k_F\sqrt{k_F^2+{M^*}^2}
             -{M^*}^3 \ln{ \left( {k_F+\sqrt{k_F^2+{M^*}^2}}
                           \over M^* \right) } \right]
\nonumber \\
             -{M^*}^3 \ln{ \left( M^* \over M\right) }
             +M^2(M^*-M)
             +{5\over 2}M (M^*-M)^2
             +{11\over 6} (M^*-M)^3.
\label{F}
\end{eqnarray}
Eq. (\ref{F}) shows there is a discontinuity in $g_s^2$ changing
from positive infinity to a negative infinity at some $k_F$ larger
than $k_0$.
This discontinuity results in the change of $g_s$ from real to a
pure imaginary and from attractive n-n scalar coupling to a 
repulsive n-n scalar coupling. To describe a long range nuclear
attraction and a short range nuclear repulsion within the 
Walecka's $\sigma$-$\omega$ model, the coupling constants
$g_s$ and $g_v$ must be real. If we assume that the same
property of nuclear interaction, i.e., the attractive scalar
interaction and the repulsive vector interaction holds at high
density too, then $g_s$ and $g_v$ must stay real.
As far as $g_s$ is real $F(k_F,M^*)$ 
should be positive as can be seen in Eq. (\ref{GS}). 
One can limit $M^*(k_F)$ such 
that $F(k_F,M^*) > 0$ by obtaining the solution of $F(k_F,M^*)=0$
for $M^*(k_F)$ which plays a role for the boundary line.
The boundary line is shown as the dash-double-dot line in Fig. 1.
The boundary line is denoted by $M'(k_F)$ as a function of $k_F$.
The region below the boundary line $M'(k_F)$ in Fig. 1 is 
forbidden for the nucleon effective mass.
Eq. (\ref{AAA}) can be integrated numerically to obtain $E_A(k_F)$ 
if $M^*(k_F)$ is known.

We choose the form of $M^*$ as a function of $k_F$ in the region 
above the boundary line $M'(k_F)$ in Fig. 1:
\begin{equation}
\label{mass}
M^*(k_F)= \left \{
\begin{array}{ll}
{ M \over   1 + {{M-M^*(k_0)}\over M^*(k_0)}
                         ({k_F \over k_0})^\alpha } 
                           ,\;\; k_F < k_0 \nonumber \\
        {{1+\beta}\over{1+\beta \left( k_F\over k_0 \right)^\delta}}
        \left[M^*(k_0)-M'(k_F)\right] + M'(k_F)
                           ,\;\; k_F > k_0
\end{array}\right.
\end{equation}
where $\alpha$, $\beta$ and $\delta$ are related by 
\begin{eqnarray}
\delta &=& { 1 \over M^*(k_0)-M'(k_0) }
           \left[ 
                 { { 2M^*(k_0) \{ M-M^*(k_0) \} \alpha } \over M }
                + 2k_0{ \left. {{dM'}\over dk_F} \right| }_{k_F=k_0}
           \right]
\nonumber \\
&+& {{ \{ 2M^*(k_0)-M \} \alpha}\over M}
\\
\beta &=& { {M^*(k_0)\{ M-M^*(k_0)\}\alpha} \over 
         { M \{M^*(k_0)-M'(k_0)\} \delta}
        - {M^*(k_0)\{M-M^*(k_0)\} \alpha}}
\end{eqnarray}
where we choose $\alpha>1$ such that $dM^*/dk_F=0$ at $k_F=0$.
We note here that $dM^*/dk_F = -\infty$ if $\alpha<1$.
As discussed above we have chosen the $M^*$ curve in the region
above the boundary line $M'(k_F)$ in the $M^*-k_F$ plane to confine
our calculations to real $g_s$. Eq. (\ref{mass}) for $M^*(k_F)$
leads to the positive effective scalar coupling constants 
$g_s^2/m_s^2$ for all $\alpha > 1.0$ as is required. Since
$g_v$ must also be real, the region of the parameter
$\alpha$ ($\alpha \gtrsim 3.5$) that leads to negative 
$g_v^2/m_v^2$, as will be shown,
should be discarded. Therefore, for the possible $M^*$ curve we have
chosen Eq. (\ref{mass}) with $1.0 < \alpha < 3.5$.
The other parameters $\beta$ and $\delta$ have been related to 
$\alpha$ in such a way that $M^*$ curve is differentiable 
at $k_F=k_0$ up to the second order.
Eq. (\ref{mass}) has been chosen such that $M^*(0)=M$ and
$M^*(k_0)$ leads to the empirical saturation point
( $k_0 = 1.42 \rm ~fm^{-1}$ and $E_A=-15.75 \rm ~MeV$ ). The $M^*$
curves are shown for each value of $\alpha$ in Fig. 1.

In numerical calculations, $M^*$ given by Eq. (\ref{mass}) for a
fixed $\alpha$ is substituted in Eq. (\ref{AAA}) and the integration
with respect to $k_F$ is performed to obtain the energy per nucleon
such that the $E_A$ curve passes through the empirical saturation
point. The nucleon binding energy curves obtained with $M^*(k_F)$
for different $\alpha$'s are plotted in Fig. 2. The calculated
incompressibility of nuclear matter increases with increasing
$\alpha$ ($K^{-1} = 61,~313$ and $565 \rm ~MeV$ for 
$\alpha = 1.0,~2.25$ and $3.5$, respectively).

Once $M^*(k_F)$ is given by Eq. (\ref{mass}) for a parameter
$\alpha$, the scalar effective coupling constant $g_s^2/m_s^2$
can be obtained from Eq. (\ref{GS}). The $g_s^2/m_s^2$ curves
for different $\alpha$'s are shown in Fig. 3.
The effective coupling constant for the vector meson can be
expressed in terms of $k_F$, $E_A$ and $M^*$ by using
Eqs. (\ref{VP}) and (\ref{EEE}).
The calculated curves for $g_v^2/m_v^2$ are depicted in Fig. 4.
The negative value of $g_v^2$ at low densities for $\alpha>3.5$
gives the upper bound of $\alpha$.

Using the density-dependent coupling constants we obtained the
equations of state of neutron matter as shown in Fig. 5, where
these are compared with the equation of state of neutron matter
with constant coupling constants\cite{Chin77}. 
The saturation
density and incompressibility of neutron matter increase with
the increasing $\alpha$.
The calculated results show that neutron matter is bound
in the regions of $1<\alpha<1.65$ and $2.79<\alpha<3.5$.
For small $\alpha$, the scalar coupling constant $g_s^2/m_s^2$ and
the vector coupling constant $g_v^2/m_v^2$ are much stronger at low
density and weaker at high $k_F$ than the fixed coupling constants.
Thus the neutron matter is bound weakly with small $k_F$ for small 
$\alpha$.
On the other hand, for large $\alpha$, both the scalar and vector 
couplings are weaker than the fixed coupling constants at low density
(small $k_F$) and the vector coupling constant $g_v^2/m_v^2$ is the
similar order as the fixed coupling constant at high density while 
the scalar coupling constant $g_s^2/m_s^2$ becomes much stronger 
than the fixed coupling constant at high density. 
This density dependence of the coupling constants for large $\alpha$ 
leads the neutron matter to be strongly bound (11.1 MeV binding 
energy per neutron).
In the mid $\alpha$ value region, the neutron matter is unbound.
However, in the region near $\alpha = 2.3$, a neutron matter has 
a resonance state at $k_F = k_0 = 1.42$ fm$^{-1}$ because the coupling
constants ($g_s^2/m_s^2$, $g_v^2/m_v^2$) for all the cases 
($\alpha$'s) are required to be the same at the density.

\section{CONCLUSION}

The results (Figs. 1 -- 4) show that,
at low density, the density-dependent coupling constants are
very sensitive to the choice of the effective mass.
However, the nuclear equation of state $E_A(k_F)$ is quite
insensitive to the coupling constants at low density.
This behavior of insensitivity reflects the fact that
the inter-nucleon separation is large at low density.
Due to the requirement of the empirical saturation in our
calculation, all the choices of the effective mass, of course,
give the same energy and coupling constants at the saturation point.
However the nuclear compressibilities are all different.

We can see from the results that the equation of state
is quite sensitive to the coupling constants at high
density as we expected.
In addition, the results show that the equation of state
of a nuclear system is also sensitive to the values of the
coupling constants at the intermediate density,
around half the saturation density.
This behavior suggests us that the detailed study of surface
vibration modes of finite nuclei is very important in extrapolating
the nuclear equation of state to a high density.
To pursue a research in this direction, we need futher investigation
of our starting point of calculation whereas here we started with 
the effective mass functions $M^*(k_F)$ just for simplicity. 
The minimum requirement in this search is that all the fits should 
pass through at $k_F=0$ and $k_F=k_0$,
i.e., a zero density limit and a nuclear saturation point.

Fig. 5 also shows that the density-dependent coupling constants
allow the existence of neutron matter within RHA.
The equations of state of neutron matter with the density-dependent
coupling constants show saturation points with different 
incompressibilities of neutron matter.
The saturation points for different $\alpha$'s are compared with
that for constant coupling constants \cite{Chin77} in Fig. 5.
The saturation state appears at higher density and 
the incompressibility is higher if $\alpha$ takes on the larger
value. 
We note here that both the resonance states with positive saturation
energy and the bound states with negative saturation energy
appear in the neutron matter calculations with the 
density-dependent coupling constants.
The saturation energies are positive in the coupling constant
region $1.65<\alpha<2.79$ where RHA does not allow a stable neutron
star.
However, for the regions $1.0<\alpha<1.65$ and $2.79<\alpha<3.5$
RHA provides bound states of neutron matter, and for $\alpha=3.5$
the binding energy appears to be $11.1~\rm MeV$
the incompressibility to be $1258~\rm MeV$
at $k_F=1.94~\rm fm^{-1}$.

This work was supported by the Ministry of Education of 
Korea through Basic Science Research Institute Grant No. 95-2422.
One of the authors (S.J.L.) would also like to acknowledge the Korea 
Science and Engineering Foundation for financial support 
under Grant No. 931-0200-035-2.

\begin{figure}
\caption{ \label{curvemass}
Nucleon effective masses $M^*$ as functions of Fermi momenta
$k_F$ for nuclear matter. The $M^*$ curves are chosen as
Eq. (\protect\ref{mass}) with a parameter $\alpha$ such that 
they lead to the empirical saturation point in nuclear matter.}
\end{figure}

\begin{figure}
\caption{Binding energies per nucleon in nuclear matter
calculated from the nucleon effective masses given 
by Eq. (\protect\ref{mass}).}
\end{figure}

\begin{figure}
\caption{Density-dependent effective coupling constant curves
for the $\sigma$-meson exchange.
The constants $g_s^2/m_s^2$ are obtained by using
Eq. (\protect\ref{GS}) with the nucleon effective masses given by
Eq. (\protect\ref{mass}).}
\end{figure}

\begin{figure}
\caption{ Density-dependent effective coupling constant curves
for the $\omega$-meson exchange. \quad}
\end{figure}

\begin{figure}
\caption{ Binding energies per nucleon for neutron matter
($\gamma=2$) calculated from the density-dependent coupling constants
$g_s^2/m_s^2$ in Fig. 3 and $g_v^2/m_v^2$ in Fig. 4.
The maximum binding energy per neutron is $11.1~\rm MeV$ for 
$\alpha = 3.5$.}
\end{figure}

\end{document}